\documentclass[twocolumn]{jpsj2} 

\title{Intermediate Phase IV in Structural Phase Transitions of TlCoCl$_3$}
\def\runauthor{}

\author{
Yoichi NISHIWAKI\thanks{E-mail address: nisiwaki@research.twmu.ac.jp}, 
Akira OOSAWA$^{1}$, Tetsuya KATO$^{2}$, 
Takumi HASEGAWA$^{3}$, Haruhiko KUROE$^{1}$, and Kazuhisa KAKURAI$^{4}$
}

\inst{%
Department of Physics, Tokyo Women's Medical University, Shinjuku-ku, Tokyo 162-8666\\
$^{1}$Department of Physics, Sophia University, Chiyoda-ku, Tokyo 102-8554\\
$^{2}$Faculty of Education, Chiba University, Inage-ku, Chiba 273-8522\\
$^{3}$Graduate School of Integrated Arts \& Sciences, Hiroshima University, Higashi-Hiroshima, Hiroshima 739-8521\\
$^{4}$Quantum Beam Science Directorate, Japan Atomic Energy Agency, Tokai, Ibaraki 319-1195
}

\recdate{\today}

\abst{%
The structural changes of TlCoCl$_3$ around phase IV (68~K $<T<$ 75~K) were studied by single-crystal neutron diffraction measurement. 
The present investigation revealed that a small but definite Bragg peak at ($\frac{1}{6}\ \frac{1}{6}\ 2$) appeared only in phase IV. 
A precursor phenomenon of the phase V ($T<$ 68~K) structure in the phase IV temperatures was also identified as the  coexistence of a slightly broadened ($\frac{1}{4}\ \frac{1}{4}\ 2$) Bragg peak. 
We confirmed that intermediate phase IV is a stable, nontransient, single phase with $2\sqrt{3}a\times 2\sqrt{3}a\times c$ unit cells. 
A possible sequence of structural phase transitions was proposed. 
}

\kword{%
TlCoCl$_3$, triangular lattice, 
neutron diffraction, structural phase transition, 
}

\begin{document}
\maketitle
\section{Introduction} 
\begin{figure}[t]
\begin{center}
\includegraphics[width=8cm]{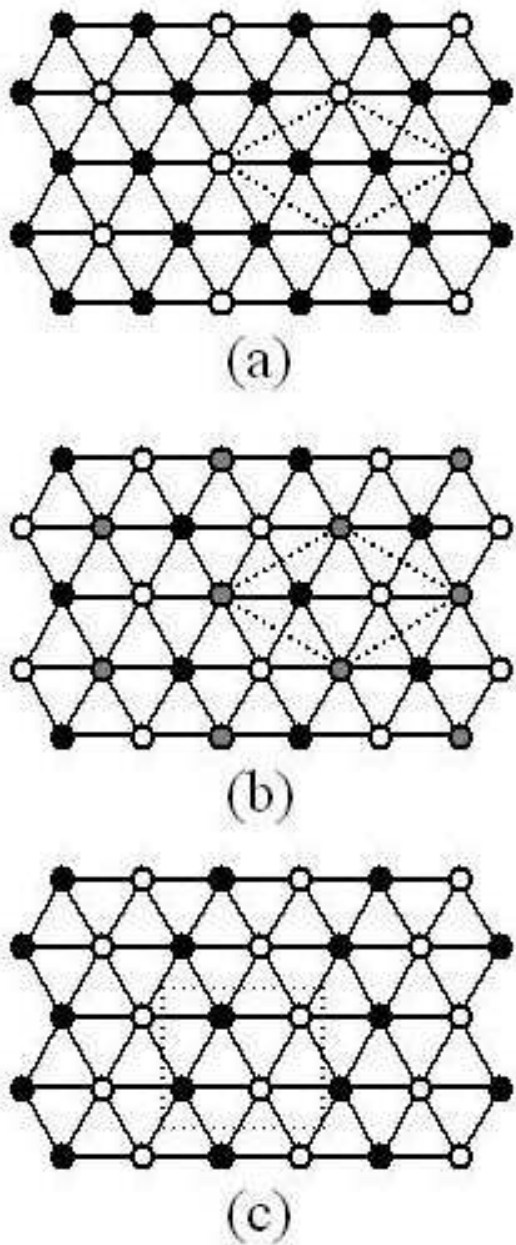}
\end{center}
\caption{Schematic drawings of -$BX_3$- chain shifts of (a) $P6_3cm$, (b) $P\bar{3}c1$, and (c) $Pbca$ structures. The chains are projected on the $c$-plane and represented by black and white circles for upward and downward chain shifts, respectively. Grey circles represent that the chain does not shift. Broken lines show the unit cell.
} 
 \label{fig:chain.eps}
\end{figure}

Among magnetic compounds with simple component ions, those with a CsNiCl$_3$-type 
structure\cite{Goodenough, Longo, Plumer1988} (space group hexagonal $P6_3/mmc$) exhibit strong one-dimensional magnetic correlations.\cite{Achiwa, PetrenkoCollins} 
For those with the $ABX_3$ chemical formula, where $A$, $B$, and $X$ are 
alkali, 3-d, and halogen ions, respectively, the linear-chain -$BX_3$- substructure along the $c$-axis 
is magnetically well separated by intervening $A$ ions. 
From the higher-temperature phase with this structure, a group of compounds 
show successive structural phase transitions. 
We call these compounds members of the KNiCl$_3$ family.\cite{Morishita, Mitsui, Yamanaka}  
Lattice distortions of KNiCl$_3$ family compounds are characterized by -$BX_3$- chain shifts along the $c$-axis.\cite{Hendrikse, Hasegawa} 
Representatives of these compounds are KNiCl$_3$\cite{Visser, VisserSolidi, Visserhigh, Petrenko96, PetrenkoCondens, Machida, Machida1997} 
and RbMnBr$_3$\cite{FinkSeifert, Kato, Kato1992, Kato2002, Heller}, 
whose room-temperature (RT) structure has triplicated 
($\sqrt{3}a\times\sqrt{3}a\times c$; $a$ and $c$ are  lattice constants of the prototype CsNiCl$_3$-structure) unit cells with chain shifts 
along the $c$-axis in an up-up-down manner, as shown in Fig. \ref{fig:chain.eps}(a). 
This phase is called phase III and the structure, sometimes called the `RT-KNiCl$_3$-structure', 
was reported to take symmetries of the space group $P6_3cm$. 
However, it was not excluded by diffraction analysis\cite{Visser, VisserSolidi} 
that the phase III structure may be another one 
with $P\bar{3}c1$ symmetry characterized 
by up-0-down type chain shifts, as shown in Fig. \ref{fig:chain.eps}(b).\cite{RbVBr3diffraction, Seifert, Kirklin} 
Interestingly, the lowest-temperature (phase V) structure is an orthorhombic 
(space group $Pbca$), quadruplicated ($2\sqrt{3}a\times\sqrt{3}a\times c$) one 
with up-up-down-down type chain shifts, as shown in Fig. \ref{fig:chain.eps}(c).\cite{Kato2002} 
The same sequence of successive structural phase transitions were hitherto 
observed in TlCoCl$_3$,\cite{Nishiwaki} TlFeCl$_3$\cite{Yamanaka2002, Zodkevitz} and RbMgBr$_3$\cite{Gesi}. 
The characteristics of higher-temperature phases I and II are not so simple that one 
should refer to the previous studies. \cite{Machida, Kato, Visserhigh} 

The reason the -$BX_3$-chain substructure is preserved and 
only the reassemblage occurs at low-temperature structural change is not clarified. 
It must be related to ferroelectricity, observed as the small spontaneous 
polarization along the chain direction in phase III with a unique temperature dependence, and also in phase IV with a notable  enhancement.\cite{Machida, Kato} 
The polarization, however, disappeared in phase V where chain shifts 
are needed for complete reassembly. 
Among the above crystals, TlCoCl$_3$ is suitable for the investigation 
of the structural change among phases III, IV, and V avoiding thermal fluctuation, 
because the transition points $T_{\rm st3}= 75$~K (III$\to$IV) and 
$T_{\rm st4}= 68$~K (IV$\to$V) are the lowest.  
 
In our previous neutron scattering study of TlCoCl$_3$\cite{TlCoCl3neutron} 
(we call it Paper I), mainly concerned with TlCoCl$_3$ magnetism, 
we confirmed the phase III and phase V structures of this crystal. 
In the temperature dependence of single-crystal Bragg peak height 
(shown as Fig. 6(b) in Paper I\cite{TlCoCl3neutron}), 
the alternation between the peaks of phases III and V were represented 
around phase IV ($T_{\rm st4} < T < T_{\rm st3}$). 
It was quite unusual that both the Bragg peaks for phase V at ($\frac14$ $\frac14$ 2) 
and for phase III at ($\frac13$ $\frac13$ 2) coexist 
at phase IV temperatures, even though the lower phase transition 
at $T_{\rm st4}$ is of the first order. 
It could become uncertain that the `phase IV' would be unstable 
or the data around those temperatures could be obtained only transiently. 
In this study, we performed single-crystal neutron scattering, 
specifically focused at these temperatures and found a Bragg peak 
at ($\frac16$ $\frac16$ 2) that appeared only in phase IV. 
This indicates that the intermediate phase IV is a stable, nontransient phase. 

\section{Experimental Procedure}
The method of preparing single crystals of TlCoCl$_3$ is stated in our previous 
report.~\cite{Nishiwaki} 
The neutron scattering for single-crystal measurement was performed 
with the triple-axis spectrometer TAS2 installed at JRR-3M of 
the Japan Atomic Energy Agency (JAEA), Tokai. 
A pyrolytic graphite ($0\ 0\ 2$) reflection was used 
as the monochromator and analyzer. 
Higher-order neutrons were removed using a pyrolytic graphite filter set 
in front of and behind the sample. 
Neutron energy was fixed at 14.7~meV ($k$ = 2.66~rad/\AA ). 
Horizontal collimations were chosen to be 
14$^{\prime}$ - 80$^{\prime}$ - 80$^{\prime}$ - 80$^{\prime}$. 
The sample, about 0.5~cm$^3$ in volume, was mounted in a 10~K refrigerator 
and the temperature of the sample was measured with a calibrated silicon diode thermometer. 
The experimental results are indexed on the prototype CsNiCl$_3$-type lattice, 
 the same as in our previous study. 
The [1$\bar{1}$0] axis was arranged vertically to measure scattering intensity  
in the ($h\ h\ l$) zone. 
The lattice constants at $T=60$~K were $a=6.91$~\AA \ and $c=5.98$~\AA.

\section{Results} 
\begin{figure}[t]
\begin{center}
\includegraphics[width=8cm]{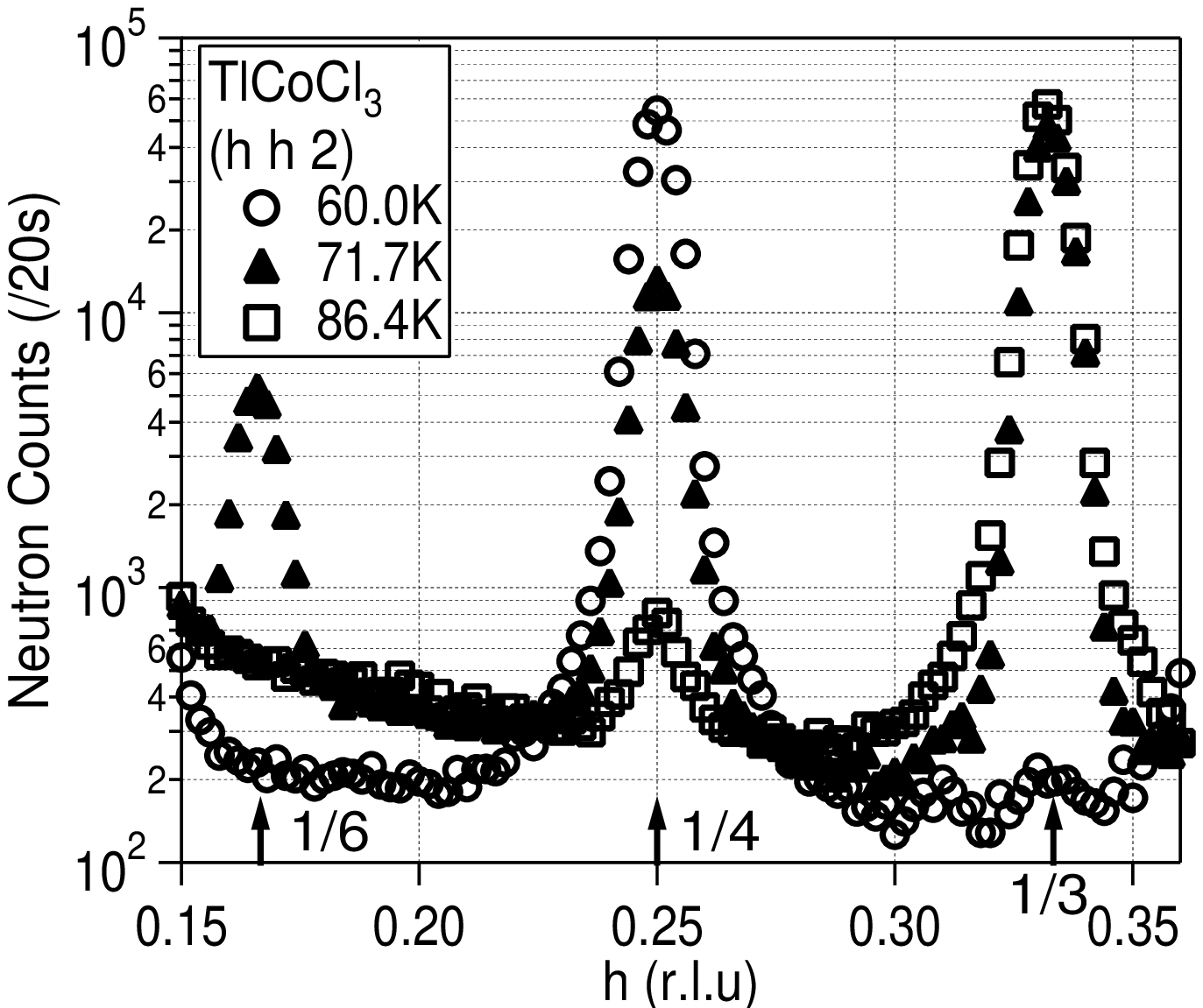}

(a)

\ 

\includegraphics[width=8cm]{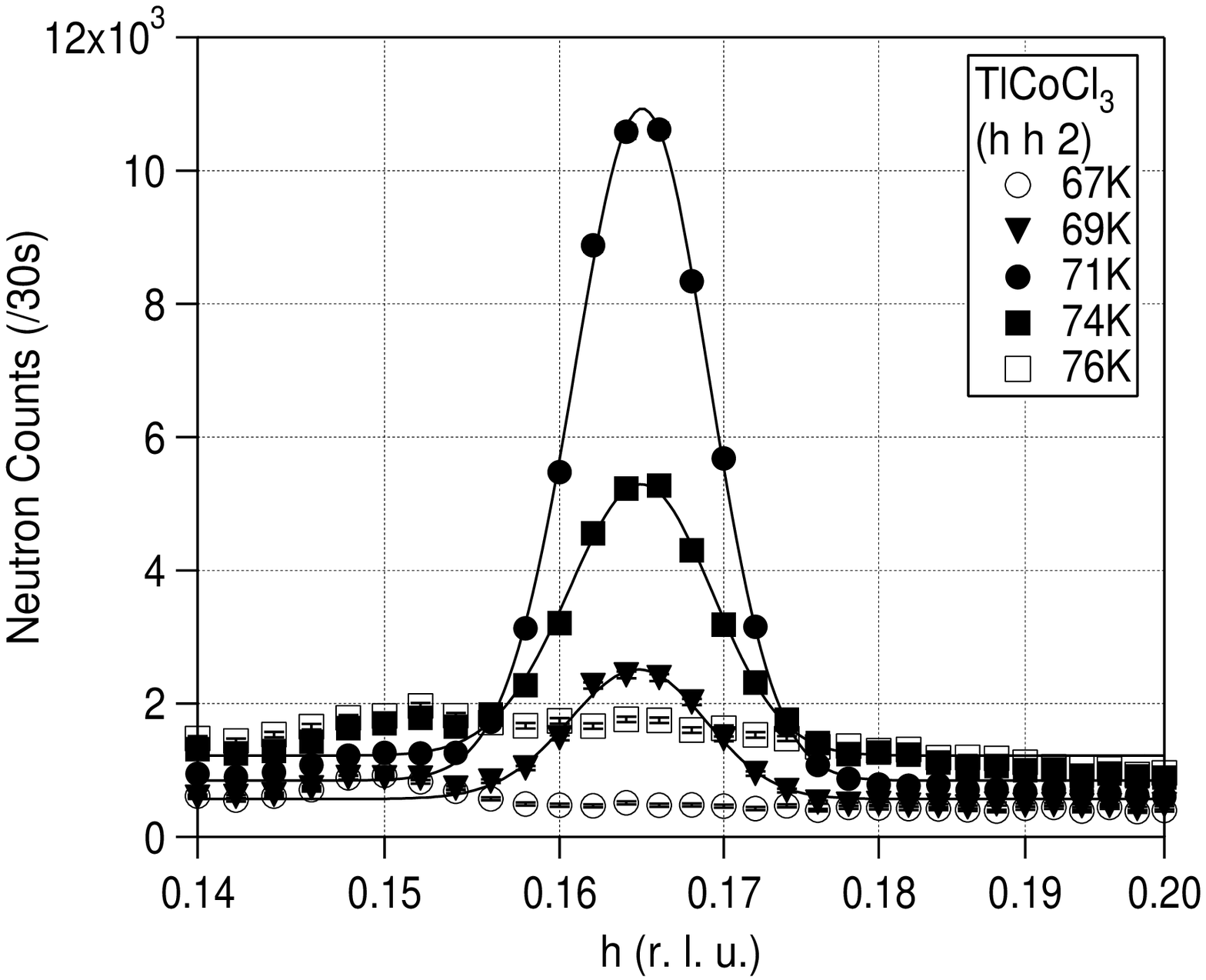}

(b)
\end{center}

\caption{
(a) Neutron diffraction intensities of single-crystal TlCoCl$_3$ 
	scanned parallel to ($h\ h\ 2$) directions at $T=$86.4~K (phase III), 
	71.7~K (phase IV) and 60.0~K (phase V).
(b) Temperature dependence of ($h\ h\ 2$) profile around ($\frac{1}{6}\ \frac{1}{6}\ 2$), measured during heating run.
}
\label{fig:Fig1}
\end{figure}

As shown in Paper I,\cite{TlCoCl3neutron} superlattice reflections 
that involve structural phase transitions are extinct in the ($h\ h\ 1$) zone 
owing to the existence of a set of symmetry operations of $\textbf{c}$ glide planes, 
one of which transforms ($x\ y\ z$) into ($y\ x\  z+\frac12$). 
This is a common nature of all structural phases of the KNiCl$_3$ family compounds. 
This is confirmed again in this study, so we start to describe 
the results of the $q$-scan profiles of ($h\ h\ 2$). 
The representative profiles from $h=$0.15 to 0.36 at 86.4~K (phase III), 
71.7~K (phase IV), and 60.0~K (phase V) are shown in Fig.~\ref{fig:Fig1}(a).  
In phase III, Bragg reflections were measured at $\frac13$ and $\frac23$, 
the latter of which appeared in a wide $q$-scan measurement. 
In phase V, in-plane reflections at $h=\frac14$ and $\frac34$ 
were observed, in addition to off-plane reflections 
at $h=\frac38$ and $\frac58$ that were detected owing to the coarse vertical 
resolution.\cite{Heller,Kato2002} 
Very small Bragg reflections at $h=\frac12$ were observed in phase V. 
These results were consistent with those in Paper I.\cite{TlCoCl3neutron} 
In phase IV, in addition to the peaks observed in phases III and V, 
 reflections at $h=\frac16$ and $\frac56$ were observed. 
The ($\frac16\ \frac16\ 2$) peak was observed only at temperatures 
between $T_{\rm st4}$ and $T_{\rm st3}$, as shown in Fig. \ref{fig:Fig1}(b). 
The full width at half maximum (FWHM) of this peak was constant in phase IV, 
at approximately 7.0$\times 10^{-3}$ \AA $^{-1}$ with a resolution limit. 
An asymmetric broad background was also observed with increasing intensity  
with increasing temperature. 

\begin{figure}[t]
\begin{center}
\includegraphics[width=8cm]{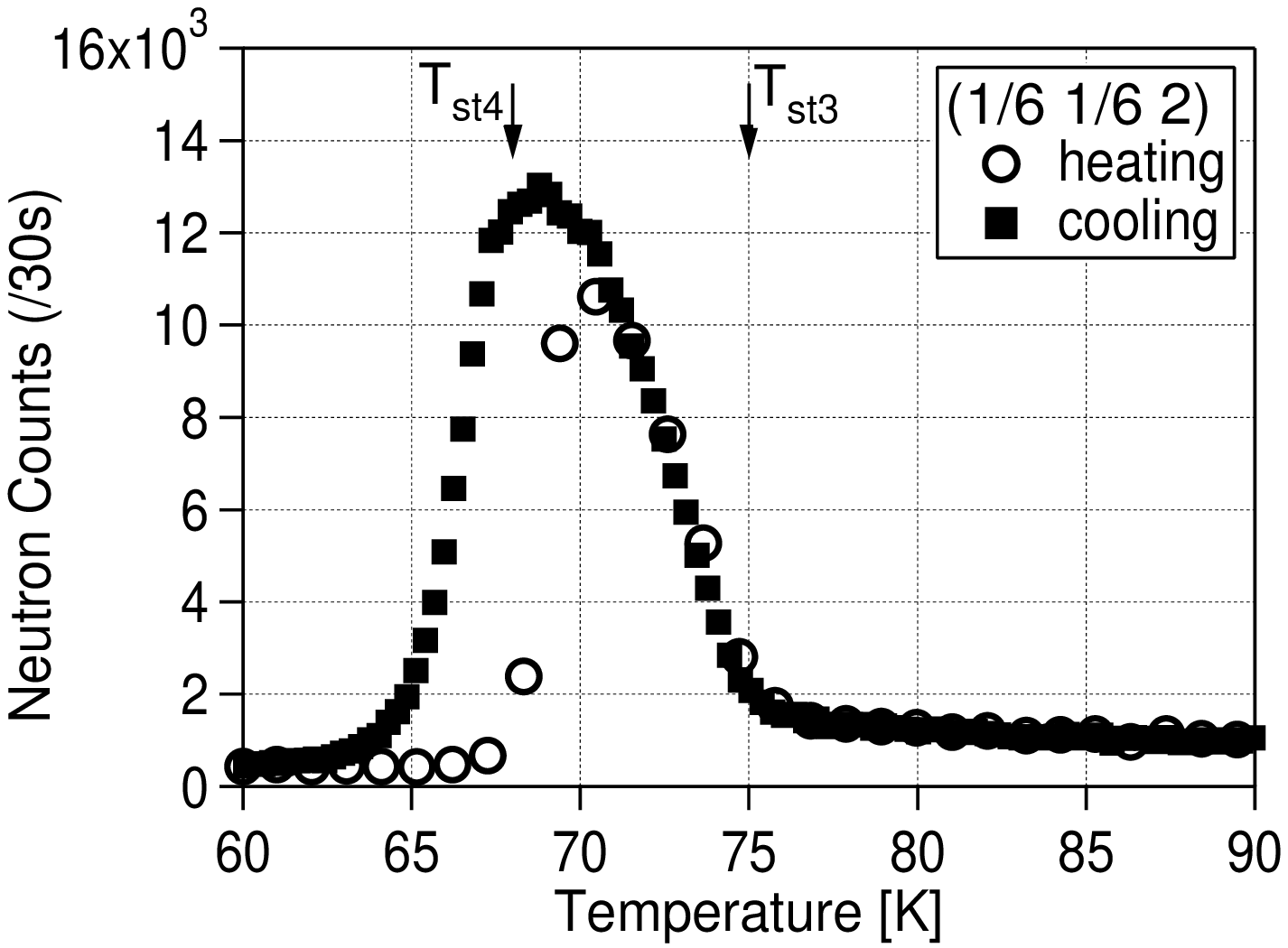}

(a)

\ 

\includegraphics[width=8cm]{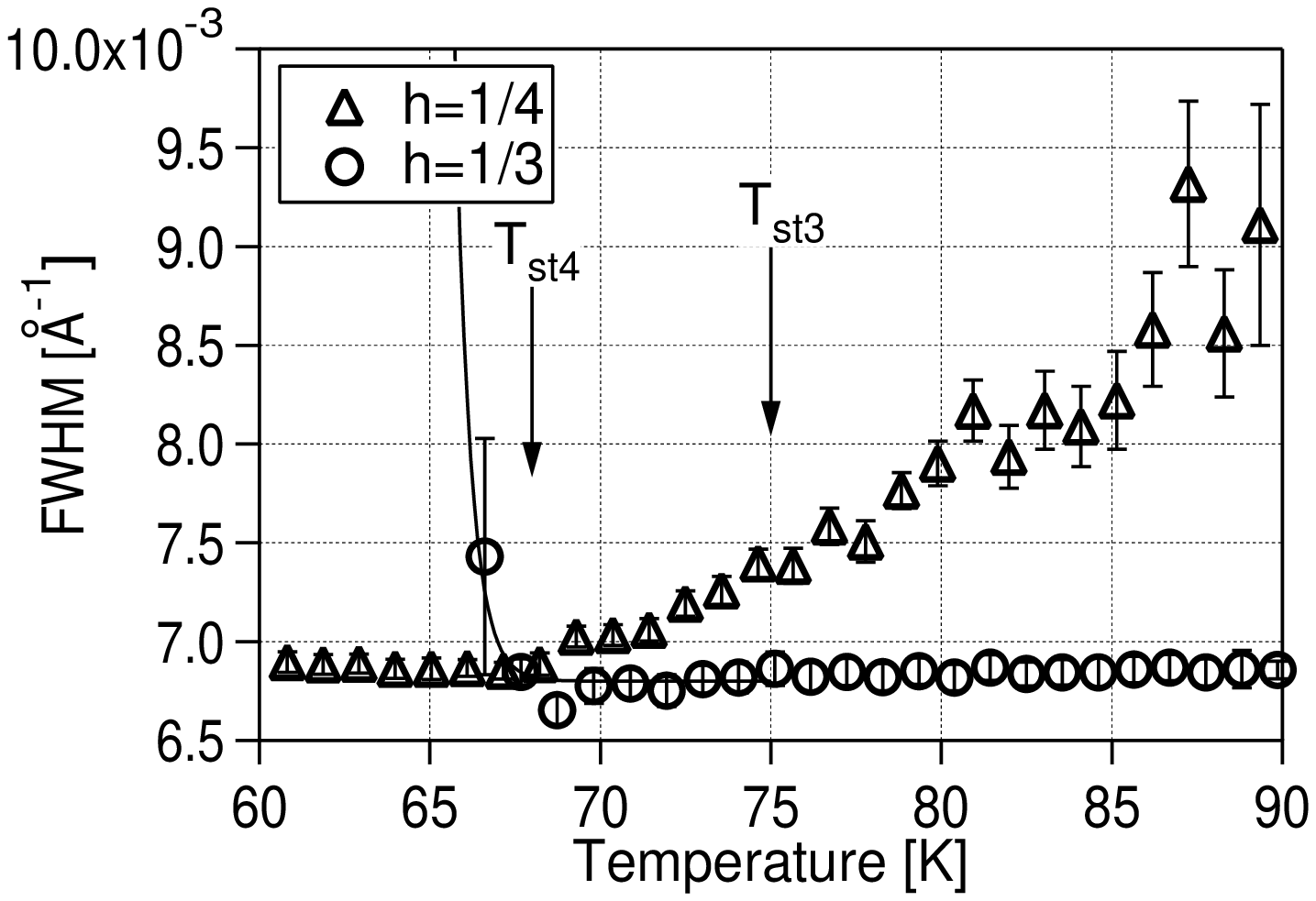}

(b)
\end{center}
\caption{
	(a) Temperature dependence of ($\frac16\ \frac16\ 2$) peak height.  
	(b) Temperature dependence of full width at half maximum (FWHM) of 
	($\frac14\ \frac14\ 2$) and ($\frac13\ \frac13\ 2$), measured during heating run. 
	An eye-guide line is given for low-temperature data of $h=\frac13$.
	}
	\label{fig:Fig2}
\end{figure}

The temperature dependence of ($\frac16\ \frac16\ 2$) 
peak height is shown in Fig.~\ref{fig:Fig2}(a). 
The heating and cooling rates were 1.8 and 0.7~deg/min, respectively. 
The thermal hysteresis around the first-order phase transition at $T_{\rm st4}$ 
was clearly indicated. 
Note that spontaneous polarization had 
a quite similar temperature dependence\cite{Nishiwaki} to the present results. 
Now, it is unlikely that phase IV could be a crossover 
between phases III and V, which is stated in Paper I.\cite{TlCoCl3neutron} 
Thus, the present results exhibited phase IV as a single phase 
that is characterized by the ($\frac{1}{6}\ \frac{1}{6}\ 2$) peak. 

The temperature dependences of peak height at ($\frac13\ \frac13\ 2$) and 
($\frac14\ \frac14\ 2$) are shown in Fig.~6(b) in Paper I\cite{TlCoCl3neutron}. 
Apparently, the `phase V peak' at ($\frac14\ \frac14\ 2$) survived at temperatures 
up to approximately 90~K, namely in phase IV and even in phase III, 
although the `phase III peak' at ($\frac13\ \frac13\ 2$) vanished 
in phase V (below $T_{\rm st4}$). 
In this study, we evaluated the temperature dependence of the FWHM of the peaks. 
The results are shown in Fig. \ref{fig:Fig2}(b).
In phase IV, the FWHM of the phase III peak was the resolution limit, although 
that of the phase V peak was definitely broadened with increasing temperature. 
Thus, a precursor phenomenon or an aftereffect with a quite strong short-range correlation 
with an ($h\ k$)$=$($\frac14\ \frac14$) periodicity was suggested 
to remain above $T_{\rm st4}$ and even above $T_{\rm st3}$.


\section{Discussion}
In this study, we found the ($\frac{1}{6}\ \frac{1}{6}\ 2$) peak appeared in phase IV. 
The scattering intensities of the ($\frac{1}{3}\ \frac{1}{3}\ 2$) and ($\frac{1}{4}\ \frac{1}{4}\ 2$) peaks are similar to  the results in Paper I. 
The maximum intensity of the ($\frac{1}{6}\ \frac{1}{6}\ 2$) peak is approximately one-tenth those of the ($\frac{1}{3}\ \frac{1}{3}\ 2$) and ($\frac{1}{4}\ \frac{1}{4}\ 2$) peaks. 
The structural model in phase IV should be one in which a small periodic modulation with a wave number $(h,\ k)=(\frac{1}{6},\ \frac{1}{6})$ is added to the phase-III structure. 

The structure of phase III was described as the freezing of
the vibrational K$_4$(1,$-$1) mode of the prototype CsNiCl$_3$ ($P6_3/mmc$)
structure,\cite{Perez, Manes} 
where (1, $-$1) indicates the antiphase combination of two order parameters in the 
K$_4$ mode. 
Here, note that the $P\bar{3}c1$ symmetry is obtained from their (1, 1) 
combination, which is the alternative symmetry for phase III, or that the  
$P3c1$ symmetry from other combinations. 
For both phases IV and V, such modes should be described as T$_2$ modes, 
because the wave vector is on the T line (connected between the $\Gamma$ 
and K points) 
and the ionic displacements along the $c$-axis must be the same for all  
chains. 
The specific displacements of T$_2$ modes and their space group symmetry 
will be listed in the further work. 
Compatibility relations for symmetry modes of $P6_3/mmc$ show that both K$_4$ and 
T$_2$ modes are compatible with the A$_2u$ mode, as listed in Tables XIV-XVII in the 
paper by Ma$\tilde{\rm n}$es {\it et al.}\cite{Manes} 
Here, we picked the $B$ ion displacements along the $c$-axis, which describe 
 nondeformed chain shifts, as shown in Table \ref{table;relation}.  
Accordingly, in both phase IV and phase V lattice distortions, the A$_2u$ mode can 
coexist as a consisting amplitude. 
Nevertheless, a zero amplitude for the A$_2u$ mode is expected for the phase-V 
structure because of the centrosymmetric combination of several order 
parameters in the T$_2$ mode in phase V. 
In phase IV, a hexagonal combination of them should be assumed. 

\begin{table}[tbp]
\caption{Selected compatibility relations of space group $P6_3/mmc$.\cite{Manes}    
}
\label{table;relation}
\begin{center}
\begin{tabular}{c|ccccc}\hline
wave vector & $\Gamma$ && T && K \\ \hline
$B_z$  & A$_{2u}$ & --- & T$_2$ & --- & K$_4$ \\
 & B$_{1u}$ & --- & T$_4$ & --- & K$_2$ \\ \hline
\end{tabular}
\end{center}
\end{table}

\begin{table}[tbp]
\caption{Possible phase transition sequences around phase IV.    
}
\label{table;sequence}
\begin{center}
\begin{tabular}{c|c|c|c}\hline
 structural phase      & III & IV & V \\
 unit cell  & $\sqrt{3}\times\sqrt{3}$ & $2\sqrt{3}\times 2\sqrt{3}$ & $2\sqrt{3}\times\sqrt{3}$\\
 mode       & $K_4\ (\frac13)$ & $T_2\ (\frac16)$ & $T_2\ (\frac14)$ \\  
 \hline
(1)& $P6_3cm$ & $P6_3cm$ & $Pbca$ \\
(2)& $P6_3cm$ & $P3c1$ & $Pbca$ \\
(3)& $P\bar{3}c1$ & $P3c1$ & $Pbca$ \\ \hline
\end{tabular}
\end{center}
\end{table}


Although the phase IV symmetry was not uniquely determined as $P3c1$, 
it is one of the most plausible candidates for the following reasons. 
Here, we pay attention on both possibilities of symmetry for 
the phase III structure ($a^{\rm III}\times a^{\rm III}\times c$ unit cells) 
as $P6_3cm$ and $P\bar{3}c1$. 
Experimental results indicate the phase transition III$\to$ IV 
at $T_{\rm st4}$ should be continuous so that the phase IV symmetry 
should be a uniaxial, hexagonal or rhombohedral subgroup of phase III symmetry. 
According to the International Tables for Crystallography 
(ITC),\cite{InternationalTables} both $P6_3cm$ and $P\bar{3}c1$ symmetries 
are themselves the maximal isomorphic subgroups of index 4 ({\it i.e.}, 
$a^{\rm IV}=2a^{\rm III}$). 
As long as the translational symmetry of phase IV exists, 
$P6_3cm$ and $P\bar{3}c1$ can be candidates;  
however, the spontaneous polarization $P_s$ of phase IV indicates directly 
that the centrosymmetric $P\bar{3}c1$ is not suitable. 
Thus, the candidate symmetry was restricted within $P6_3cm$ itself and the subgroups of 
$P6_3cm$ and $P\bar{3}c1$. 
The maximal non-isomorphic subgroups of $P6_3cm$ are $P6_3$, $P3c1$ and $P31m$, 
and those of $P\bar{3}c1$ are $P321$, $P\bar{3}$ and $P3c1$, as shown in ITC.\cite{InternationalTables} 
Among these five subgroups, only $P3c1$ is suited for the observed extinction rules 
for the existence of a $\textbf{c}$ glide plane symmetry. 
Consequently, three kinds of phase transition sequence around phase IV can be proposed, as shown in Table \ref{table;sequence}.  

We assume that the sequence $P\bar{3}c1$ (phase III)$\to  P3c1$ (IV)$\to  Pbca$ (V) is the most plausible because 
the spontaneous polarization is quite small in phase III but relatively large in phase IV.\cite{Nishiwaki} 
If the phase III structure is $P6_3cm$, it can hardly explain why the spontaneous polarization in phase IV is much larger than that in phase III. 
It might be that the spontaneous polarization is zero in phase III ($P\bar{3}c1$) and only appears in phase IV ($P3c1$). 
The quite small polarization in phase III may be caused by phase-shift domains or twinning in the $P\bar{3}c1$ structure. 
The regions of up-up-down, up-up-down-down and up-0-down-0 type chain shifts tend to appear at phase boundaries 
when the phase transtion occurs from $P6_3/mmc$ to $P\bar{3}c1$.\cite{HasegawaDth} 
The up-up-down region at the boundaries can induce the small polarization. 
In addition, another up-up-down-down region on the boundaries might explain the present results 
showing the precursor or short-range phase V ordering even above $T_{\rm st4}$. 
Therefore, a structural reinvestigation of phase III using a four-axis spectrometer or some another techniques and the determination of the phase IV structure in further works are needed to confirm this $P\bar{3}c1$-structure scenario for  successive structural phase transitions. 
The phase transition sequence of TlCoCl$_3$ should be common to those of KNiCl$_3$, RbMnBr$_3$ and TlFeCl$_3$. 

In conclusion, we performed single-crystal neutron diffraction measurement in TlCoCl$_3$. 
Phase IV was shown as a single phase with $2\sqrt{3}a\times 2\sqrt{3}a\times c$ unit cells, 
because a Bragg peak at ($\frac{1}{6}\ \frac{1}{6}\ 2$) appeared only 
in phase IV. The temperature dependence of this peak is quite similar 
to that of spontaneous polarization. 
We confirmed that phase III, IV and V structures are characterized by 
the ($\frac{1}{3}\ \frac{1}{3}\ 2$), ($\frac{1}{6}\ \frac{1}{6}\ 2$) and 
($\frac{1}{4}\ \frac{1}{4}\ 2$) peaks, respectively. 
A possible structural phase transition sequence was proposed. 

\section*{Acknowledgment}

The authors would like to express their sincere thanks to K. Iio, T. Sekine and T. Asahi for useful discussions.






\end{document}